\documentclass[useAMS,usenatbib]{mn2e}

\usepackage{graphics}
\usepackage{graphicx}
\usepackage{amsxtra}
\usepackage{array,longtable}
\usepackage{journals}
\usepackage{url}
\usepackage{latexsym}
\usepackage{amssymb}

\usepackage{ulem}

\newcommand{\ser}{S\'ersic}

\title
[Pitch angle variations in spiral galaxies]
{Pitch angle variations in spiral galaxies
}
\author[S.S.~Savchenko, V.P.~Reshetnikov]
{S.S.~Savchenko$^1$\thanks{E-mail: savchenko.s.s@gmail.com}
and V.P.~Reshetnikov$^{1,2}$
\\
$^1$St.Petersburg State University, Universitetskij pr. 28, 198504
St.Petersburg, Stary Peterhof, Russia \\
$^2$Isaac Newton Institute of Chile, St Petersburg Branch
}

\begin{document}

\date{Accepted 2013 August 28. Received 2013 August 26; in original form 2013 June 25}

\pagerange{\pageref{firstpage}--\pageref{lastpage}} \pubyear{2013}

\maketitle

\label{firstpage}
\begin{abstract}

We present a detailed photometric study and measurements of spiral arm pitch angles
for a sample of 50 non-barred or weakly barred grand-design spiral galaxies 
selected from Sloan Digital Sky Survey.
In order to find pitch angles, we used a new method based on the window Fourier 
analysis of their images. This method allows us not only to infer the average 
pitch angle, but to obtain its value as a function of galactocentric radius as well. 
Our main results are as follows: 

(1) Spiral arms of most galaxies cannot be described by a single value of the 
pitch angle. About 2/3 of galaxies demonstrate pitch angle variations exceeding
20\%. In most galaxies in the sample their pitch angle decreases by increasing the distance from the centre. 

(2) Pitch angle variations correlate with the properties of galaxies --
with the shape of the surface brightness distribution (envelope-type or truncated
disc), and with the sign of stellar disc colour gradient.

(3) More luminous and bright bulges produce more tightly wound spiral arms, that is 
in agreement with current models for spiral arms formation.

\end{abstract}


\section{Introduction}

Spiral patterns are the most prominent features of disc galaxies. Spiral arm
shape (among other parameters of galaxies) changes along the Hubble
sequence. Disc galaxies of early Hubble types have tightly wounded spiral arms,
while ones of later types have more wide open spiral structure.

The degree of tightness of the spiral structure is described by the \textit{pitch angle}.
The pitch angle is the angle between the tangent to spiral arm and the perpendicular
to the radius-vector drawn from the centre of the galaxy (e.g., \citealt{binney1987}).

Usually, for obtaining the pitch angle value, spiral arm is described by
some model (for example, logarithmic, archimedean or hyperbolic spiral). 
This model makes limitations on how the value of the pitch angle can
vary with the distance from the galaxy centre. For example, in the case of
logarithmic spiral (which is the most commonly used for representation of spiral
structure) the pitch angle does not change with distance at all.
Other models, however, can show different behaviour of the pitch angle.
Thus, the value of the pitch angle of archimedean spiral decreases with the
distance, whereas for hyperbolic spiral it increases.

Real galaxies, however, can show more various behaviour of the pitch
angle than the models listed above (see discussion in \citealt{ringermacher2009}).
It explains uncertainty in selection
of the model for spiral pattern approximation: while most authors choose
logarithmic spirals (e.g., \citealt{ma2001}, \citealt{seigar2008}), some use 
archimedean (\citealt{karachentsev1967}) and hyperbolic (\citealt{kennicutt1981}) 
spirals, or even more complicated models (\citealt{ringermacher2009}).

It is possible to obtain the pitch angle value as a function of a distance 
from the galactic centre by fitting different parts of the spiral structure 
separately (e.g., \citealt{russel1992}, \citealt{russel1993})
or by using two-dimensional Fourier analysis of a spiral pattern (\citealt{s2012}).
In this case the shape of the spiral structure is
represented by a set of angles of tangents rather than by a formula with fixed
parameters, and the pitch angle value does not obey the specific model.

The correlation between the galactic central mass concentration and the pitch angle
is predicted by current models for spiral arms formation (see
detailed discussion in \citealt{grand2013}, \citealt{berr2013}). 
Recent observations suggest that there is a link between the tightness 
of spiral structure and the shear rate of differentially rotating discs
of spiral galaxies (\citealt{seigar2005}, \citealt{seigar2006}). 
Since the shear rate is determined by the mass distribution, this correlation
probably reflects a correlation between the mass of central region and spiral
arm pitch angle.

Shear rate is an indirect measurement of the central mass concentration. It is important
to check pieces of independent evidence -- for instance, does the pitch angle correlates
with the total bulge mass (or luminosity) or with other parameters of the galaxies?

In this work, we present measurements of the pitch angle values and
their radial variations in a sample of grand-design galaxies selected from 
Sloan Digital Sky Survey (SDSS). 
Special attention was paid to the comparison of the spiral pattern
characteristics with general photometric parameters of galaxies (their bulges and
discs) in order to check possible mutual correlations. 

The structure of the paper is as follows: in Section 2, we outline our
sample; in Section, 3 we describe our methods; results are presented in Section 4,
and Section 5 contains our main findings.

\section[]{The sample}

The study of the pitch angle variations requires a sample of galaxies
with a prominent, well shaped spiral structure, which can be traced for
considerable distance from the centre of the galaxy. 

The selection of objects for our sample was mainly based on the EFIGI
(Extraction de Formes Idealisées de Galaxies en Imagerie, \citealt{efigi}) 
catalogue, which contains morphological parameters of 4458 galaxies. This allows 
us to create a primary  sample by constraining some of these parameters.

An automatic estimate
of properties of the spiral structure requires galaxies with high contrast and
regular spiral pattern without significant contamination by foreground objects. Thus, we
have constrained \textit{arm strength, perturbation, flocculence} values to get better looking 
galaxies. The gravitational interaction between galaxies leads to distortion
of spiral pattern, so the \textit{multiplicity} parameter has to be equal to zero.

Another one constraining criteria for our sample is the inclination angle.
We were interested in galaxies with intermediate values of inclination
angle (about $40\degr$--$60\degr$) only. More inclined galaxies have too big distortions
due to the projection effects, whereas ones with less inclination cannot be
processed by Poltorak--Freedman Monotony of Spiral Arms  (MSA) algorithm (see below).

EFIGI parameters are rough, so the final selection was made by eye. At this
stage, we have also rejected galaxies with more than two spiral arms.

Our final sample consists of 50 galaxies with two large-scale spiral arms.
The galaxies of our sample are mostly without or with insignificant bars,
although some galaxies have prominent bars (e.g. PGC~23028 and PGC~22596).  
Table 1 summarizes the main characteristics of the galaxies. 
The columns of the table are: 

\begin{itemize}
\item[(1)]PGC -- number according to the Principal General Catalogue; 
\item[(2)]Type -- morphological type according to HyperLeda\footnote{http://leda.univ-lyon1.fr} (\citealt{paturel2003});
\item[(3)]$M_g$ -- absolute luminosity in the $g$ filter (based on apparent magnitude from SDSS and 
luminosity distance and galactic absorption from NED\footnote{http://ned.ipac.caltech.edu});
\item[(4)]$i$ -- inclination in degrees ($i=0$ for face-on galaxy), see Section 2.2;
\item[(5)]$V_{{max}}$ -- maximum rotational velocity (from HyperLeda \textit{vmaxg} parameter), corrected for our 
inclination $i$;
\item[(6)]$\mu_b^0$ -- bulge central surface brightness in \textit{g} band (this one and the 
next five parameters see in Section 2.3);
\item[(7)]$r_e$ -- effective radius of a bulge in \textit{g} band;
\item[(8)]$n$ -- \ser~ index of a bulge in \textit{g} band;
\item[(9)]$\mu_{d_1}^0$ -- central surface brightness of an inner disc in \textit{g} band;
\item[(10)]$h_1$ -- exponential scale of an inner disc in \textit{g} band;
\item[(11)]$\mu_{d_2}^0$ -- central surface brightness of an outer disc in \textit{g} band;
\item[(12)]$h_2$ -- exponential scale of an outer disc in \textit{g} band;
\item[(13)]$(B/T)_g$ -- bulge-to-total luminosity ratio in \textit{g} band; 
\item[(14)]$\langle\psi\rangle$ -- mean pitch angle in degrees in \textit{g} band. Footnotes
show values of the pitch angle measured by other authors. Although our sample has very little intersection
with samples of other authors, some galaxies have one or more published estimates of the pitch angle;
\item[(15)]$\Delta\psi/\langle\psi\rangle$ -- amplitude of pitch angle variation to mean 
pitch angle ratio in \textit{g} band (see Sect. 3).

\end{itemize}

\begin{table*}
 \centering
 \begin{minipage}{190mm}
  \parbox[t]{190mm} { \caption{The sample}}
  \begin{tabular}{rcccccccccccclc}
\hline
PGC&Type&$M_g$&$i$&$V_{max}$&$\mu_b^0$&$r_e$&$n$&$\mu_{d_1}^0$&$h_1$&$\mu_{d_2}^0$&$h_2$&$\frac{B}{T}$&$\langle\psi\rangle$&$\frac{\Delta\psi}{\langle\psi\rangle}$   \\
    &    & $^m$ & $(\degr)$ & $\frac{km}{sec}$   & $\frac{^m}{\square''}$ & $''$ & & $\frac{^m}{\square''}$ & $''$ &$\frac{^m}{\square''}$ & $''$ & & $(\degr)$ & \\

\hline
1909 & SBbc & -19.37 & 63 & 150 & --- & --- & --- & 19.42 & 7.8 & 20.14 & 10.2 & 0.00 & 17.3 & 0.69 \tabularnewline
2182 & Sbc & -20.76 & 42 & 208 & 17.38 & 2.4 & 1.62 & 21.61 & 21.6 & 19.99 & 10.2 & 0.08 & 11.9 & 0.73 \tabularnewline
10559 & SABa & -21.26 & 55 & 203 & 19.29 & 1.8 & 0.39 & 19.32 & 7.2 & 21.21 & 15.0 & 0.05 & 10.1 & 0.15 \tabularnewline
13535 & SABc & -20.29 & 57 & 137 & 19.36 & 9.0 & 1.55 & 22.42 & 37.8 & 21.20 & 17.4 & 0.15 & 12.6 & 0.40 \tabularnewline
21475 & Sc & -19.47 & 60 & 144 & 21.05 & 1.8 & 0.6 & 20.79 & 13.2 & 18.73 & 8.4 & 0.00 & 14.1 & 0.30 \tabularnewline
22596 & SBb & -20.68 & 59 & 175 & 18.12 & 1.8 & 0.59 & 20.21 & 13.2 & 19.11 & 9.6 & 0.06 & 19.4 & 0.36 \tabularnewline
23028 & Sb & -19.60 & 54 & 168 & 16.82 & 8.4 & 2.39 & --- & --- & 19.35 & 13.2 & 0.11 & $16.7^a$ & 0.32 \tabularnewline
23337 & SABa & -20.71 & 44 & 265 & 13.9 & 4.2 & 3.05 & 22.41 & 71.4 & 21.20 & 16.2 & 0.31 & 10.6 & 0.17 \tabularnewline
24423 & SBb & -19.52 & 45 & --- & 20.4 & 5.4 & 1.38 & 22.20 & 35.4 & 21.32 & 10.8 & 0.09 & 20.5 & 0.25 \tabularnewline
26528 & Sb & -20.53 & 69 & 243 & 20.96 & 1.8 & 0.26 & 20.67 & 10.2 & 18.51 & 5.4 & 0.01 & 19.2 & 0.37 \tabularnewline
27121 & Sbc & -20.94 & 47 & 235 & 18.57 & 4.2 & 1.38 & 20.69 & 8.4 & 22.27 & 16.8 & 0.25 & 10.4 & 0.11 \tabularnewline
30694 & Sbc & -20.43 & 50 & 193 & 19.73 & 3.0 & 1.38 & 21.20 & 12.0 & 18.61 & 6.0 & 0.02 & 17.1 & 0.13 \tabularnewline
31883 & Sc & -18.99 & 59 & 182 & 15.81 & 23.4 & 3.4 & 21.36 & 80.4 & 21.02 & 47.4 & 0.12 & $14.2^b$ & 0.07 \tabularnewline
31917 & SABb & -21.85 & 58 & 358 & 20.77 & 3.0 & 0.45 & 20.95 & 15.0 & 19.79 & 7.8 & 0.05 & 11.7 & 0.44 \tabularnewline
32831 & Sb & -20.46 & 54 & --- & 19.26 & 2.4 & 0.84 & 20.60 & 6.6 & 21.74 & 12.0 & 0.15 & 10.7 & 0.30 \tabularnewline
33040 & Sab & -20.95 & 34 & 227 & 12.8 & 1.8 & 3.58 & 20.41 & 7.2 & 21.22 & 11.4 & 0.13 & \hphantom{0}8.9 & 0.44 \tabularnewline
33719 & SABb & -21.07 & 54 & --- & 18.62 & 3.0 & 0.95 & 21.30 & 12.6 & 20.19 & 8.4 & 0.16 & 10.3 & 0.53 \tabularnewline
34599 & SBb & -21.46 & 49 & 295 & 21.27 & 1.8 & 0.56 & 20.73 & 10.2 & 16.89 & 3.6 & 0.00 & 17.3 & 0.22 \tabularnewline
35952 & SABb & -19.47 & 54 & 145 & 13.57 & 1.2 & 2.91 & 21.29 & 18.6 & 19.82 & 9.0 & 0.05 & 17.3 & 0.06 \tabularnewline
38024 & Sbc & -21.13 & 46 & 198 & 20.07 & 3.0 & 0.55 & 21.12 & 15.0 & 22.01 & 22.8 & 0.07 & $15.0^c$ & 0.17 \tabularnewline
38885 & SABa & -18.47 & 54 & 138 & 20.05 & 4.2 & 0.88 & 21.95 & 61.2 & 20.34 & 10.8 & 0.07 & 14.8 & 0.12 \tabularnewline
38916 & Sbc & -19.41 & 56 & 135 & 21.15 & 3.0 & 0.37 & 21.39 & 26.8 & --- & --- & 0.01 & 14.2 & 0.09 \tabularnewline
39038 & Sab & -21.45 & 56 & 222 & --- & --- & --- & 20.96 & 6.9 & --- & --- & 0.00 & 18.1 & 0.04 \tabularnewline
39479 & Sc & -20.39 & 54 & 200 & 20.46 & 4.2 & 1.03 & 21.50 & 49.2 & 20.94 & 21.0 & 0.02 & 14.3 & 0.13 \tabularnewline
39775 & Sbc & -19.80 & 60 & 186 & 20.33 & 2.4 & 0.51 & 20.55 & 13.2 & 21.74 & 21.6 & 0.03 & 15.9 & 0.32 \tabularnewline
39793 & SBc & -21.42 & 48 & 229 & 21.53 & 1.8 & 0.43 & 20.89 & 7.8 & 19.09 & 4.2 & 0.01 & 17.5 & 0.38 \tabularnewline
40030 & Sa & -19.36 & 61 & --- & 18.94 & 3.6 & 0.8 & 20.98 & 26.4 & 20.68 & 19.2 & 0.08 & 10.9 & 0.09 \tabularnewline
41244 & S0-a & -20.02 & 63 & 200 & 19.75 & 3.6 & 0.71 & 21.37 & 9.0 & 22.43 & 16.8 & 0.20 & 11.8 & 0.47 \tabularnewline
42847 & SABc & -19.67 & 45 & 133 & 20.19 & 4.2 & 0.91 & 21.18 & 26.4 & 22.20 & 38.4 & 0.03 & 13.5 & 0.07 \tabularnewline
45833 & Sb & -21.16 & 60 & --- & 18.83 & 2.4 & 0.8 & 22.06 & 23.4 & 18.35 & 5.4 & 0.07 & 11.5 & 0.31 \tabularnewline
47011 & SBbc & -21.07 & 65 & 230 & 17.48 & 6.0 & 1.83 & 21.06 & 28.8 & 19.66 & 12.6 & 0.13 & 18.9 & 0.21 \tabularnewline
47855 & SABc & -21.26 & 54 & 207 & 21.31 & 1.2 & 0.17 & 20.71 & 17.4 & 18.54 & 8.4 & 0.00 & 18.0 & 0.21 \tabularnewline
48392 & Sc & -18.77 & 55 & 123 & 21.09 & 5.4 & 0.86 & 21.54 & 26.4 & 19.65 & 15.0 & 0.01 & 21.9 & 0.54 \tabularnewline
49540 & Sa & -21.19 & 53 & 266 & 18.38 & 3.0 & 0.85 & 20.86 & 12.0 & 22.09 & 22.2 & 0.21 & 14.0 & 0.20 \tabularnewline
49555 & Sbc & -18.97 & 46 & 173 & 19.72 & 9.0 & 0.94 & 21.36 & 66.6 & 20.56 & 49.8 & 0.03 & $\hphantom{0}9.9^d$ & 0.23 \tabularnewline
50610 & SBb & -20.12 & 47 & --- & 21.06 & 1.8 & 0.3 & 20.60 & 7.8 & 22.20 & 12.6 & 0.05 & 16.1 & 0.32 \tabularnewline
50897 & Sa & -20.48 & 60 & 188 & 20.21 & 2.4 & 0.39 & 19.93 & 8.4 & 21.26 & 15.0 & 0.04 & 21.4 & 0.20 \tabularnewline
51541 & Sb & -19.75 & 59 & 175 & 19.84 & 4.8 & 1.38 & 20.70 & 15.0 & 19.52 & 10.2 & 0.03 & 17.6 & 0.57 \tabularnewline
51733 & Sbc & -19.26 & 51 & 121 & 20.89 & 4.8 & 0.6 & 21.98 & 25.2 & 20.78 & 9.6 & 0.11 & 16.4 & 0.26 \tabularnewline
54018 & SABc & -19.97 & 55 & 170 & 19.74 & 4.8 & 1.29 & 21.86 & 73.2 & 19.79 & 15.6 & 0.02 & $17.6^e$ & 0.27 \tabularnewline
54200 & SBbc & -21.33 & 51 & --- & 21.88 & 1.8 & 0.29 & 21.35 & 18.6 & 20.59 & 7.8 & 0.01 & 24.5 & 0.30 \tabularnewline
54232 & SABb & -20.93 & 67 & 654: & 20.83 & 3.6 & 0.47 & 21.73 & 43.8 & 19.84 & 9.6 & 0.04 & 11.0 & 0.10 \tabularnewline
55213 & SBbc & -21.19 & 43 & 245 & 18.29 & 3.0 & 1.37 & 21.89 & 13.8 & 19.56 & 6.0 & 0.12 & 17.4 & 0.14 \tabularnewline
55792 & SABb & -21.55 & 36 & 355 & 20.21 & 3.0 & 0.64 & 22.58 & 42.6 & 20.74 & 8.4 & 0.09 & 10.6 & 0.15 \tabularnewline
55601 & SABb & -20.81 & 50 & 175 & 18.88 & 3.0 & 0.92 & 20.87 & 12.6 & 21.73 & 17.4 & 0.15 & 12.3 & 0.55 \tabularnewline
57800 & SABb & -21.11 & 55 & --- & 21.09 & 2.4 & 0.1 & 20.78 & 10.2 & 20.33 & 8.4 & 0.03 & 13.8 & 0.12 \tabularnewline
58596 & Sb & -21.57 & 58 & 295 & 17.9 & 4.8 & 1.49 & 22.30 & 114.6 & 20.62 & 12.6 & 0.21 & 14.6 & 0.15 \tabularnewline
59222 & Sb & -21.06 & 45 & 213 & 19.83 & 2.4 & 0.95 & 22.49 & 16.2 & 21.13 & 7.2 & 0.13 & 14.4 & 0.13 \tabularnewline
65310 & SABb & -20.75 & 51 & --- & 19.03 & 3.6 & 1.86 & 21.45 & 16.2 & 18.72 & 7.2 & 0.01 & 17.9 & 0.42 \tabularnewline
1346399 & S? & -20.71 & 53 & 193 & 21.13 & 3.0 & 0.56 & 21.09 & 5.4 & 23.28 & 14.4 & 0.14 & 21.5 & 0.91 \tabularnewline
\hline
\multicolumn{10}{l}{$^a$ -- Kennicutt (1981): $13\degr$, Ma (2001): $12.2\degr$;} \tabularnewline
\multicolumn{10}{l}{$^b$ -- Kennicutt (1981): $13\degr$, Seigar et al. (2006): $13.6\degr$, Ma et al. (1998): $14.2\degr$;} \tabularnewline
\multicolumn{10}{l}{$^c$ -- Ma (2001): $13.8\degr$, Ma et al. (1998): $17.7\degr$;} \tabularnewline
\multicolumn{10}{l}{$^d$ -- Kennicutt (1981): $9\degr$, Ma et al. (1998): $8.4\degr$;} \tabularnewline
\multicolumn{10}{l}{$^e$ -- Ma (2001): $15.8\degr$.} \tabularnewline
\end{tabular}
\end{minipage}
\end{table*}

\begin{figure}
\centering
\includegraphics[width=7.5cm, angle=0, clip=]{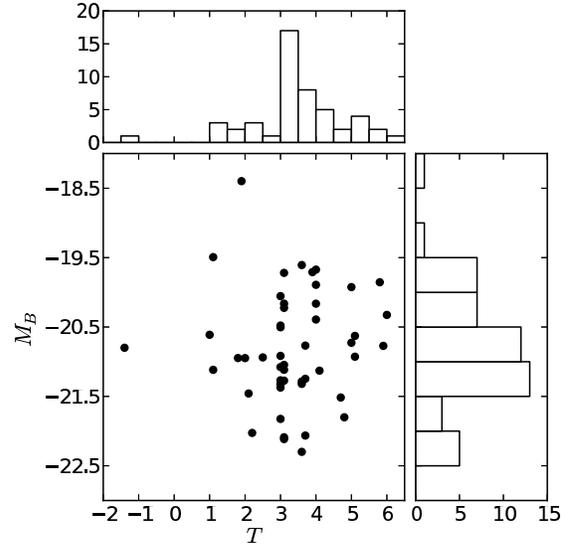}
\caption{Distribution of the galaxies of the our final sample by
luminosities and morphological types. }
\label{fig1}
\end{figure}

Fig. \ref{fig1} represents the distribution of our sample galaxies by
luminosities and morphological types (according to HyperLeda). As one can see,
spiral galaxies of all morphological types are presented in the sample, with some
excess of Sb--Sbc types.

g- and r-band images for all galaxies in our sample were obtained from
SDSS DR9\footnote{http://www.sdss3.org/dr9/} data base. We have used the
 SWARP program (\citealt{bertin2002}) to concatenate images of galaxies,
splitted on two or more nearby fields.

\section[]{Methods}

Analysis of a galaxy image consists of three main steps. The first step consists of the
determination a galaxy orientation parameters and correction
of the image for projection effects. The second one consists of the
decomposition of a galaxy on to bulge and disc subcomponents and their subtraction 
from the original image. The final step is the pitch angle measurement itself.
In this section we describe these steps one by one.

\subsection{Inclination}

The projection of a galaxy on the celestial sphere distorts its visible shape and
the degree of this distortion depends on value of an inclination of the galaxy.
A study of a galaxy spiral structure, therefore, requires measurements of
its orientation which can be described by two parameters: \textit{an inclination
angle} of a galaxy plane to the line of sight and \textit{a position angle} of
galaxy major axis.

In this work we have used the MSA method 
(\citealt{poltorak2007}, \citealt{fp2010}).
This method allows us to obtain values of both inclination and position angle.

It is possible to represent a spiral arm of a galaxy as a function $r=r(\Theta)$ (in polar
coordinate system). The main idea of the MSA method is based on the assumption
that $\frac{dr}{d\Theta} > 0$ (i.e. the spiral arm is a monotony function).

This assumption however may be violated due to projection effect.
Indeed, the projection on the celestial sphere decreases the visible size of the galaxy in 
one direction (along minor axis) and does not change it in perpendicular one (major axis).
Thereby some parts of the spiral arm can show non-monotony behaviour.

One can deproject the image of the galaxy for all possible values of inclination
and position angle (formally $0\degr\le i \le 90\degr$, $0\degr\le PA < 180\degr$)
and find those which give monotony spiral structure.

Accuracy of this method depends on a true value of inclination: for a less
inclined galaxies it gives a bigger uncertainty. The lower limit of the method
is about 30$\degr$ -- 40$\degr$.

This method works well for galaxies with extended and prominent spiral arms, such as
in our sample. Widely used method based on measurements of apparent ellipticity
of outer isophotes can introduce significant errors due to distortion of isophotes shape by bright
spiral arms, projected stars etc. In our previous work (\citealt{sr2011}) we have 
compared the MSA and isophotes methods and have found good mutual agreement. However, for several 
galaxies, the apparent isophote flattening method gives, obviously, incorrect values due 
to various kinds of peculiarities in the shapes of faint outer isophotes.

We have applied the MSA algorithm to all galaxies in our sample; the obtained
values are listed in Table~1. The maximum rotational velocities in 
Table~1 ($V_{max}$) were obtained from the HyperLeda values corrected for our 
inclination estimates.

\subsection{Bulge--disc decomposition}

The pitch angle of spiral arms changes systematically along the Hubble
sequence as well as many other morphological parameters (bulge-to-total ratio,
\ser~ index of a bulge, etc.). One may expect the existence of correlations
between the pitch angle value and various parameters of galaxies. For 
searching such correlations, we have made the decomposition of galaxies surface
brightness distributions on to bulge and disc components.

We have used azimuthal averaging to get one-dimensional photometric profile 
of a galaxy $\mu=\mu(r)$.
The averaging was made along ellipsis which ellipticity and position angle
are correspond to the inclination and the position angle of the galaxy. To minimize
the influence of background objects, we applied the iterative $3\sigma$
rejection procedure before averaging.

Bulges of galaxies are usually described by \ser~ profiles, whereas discs
can be well represented  by a simple exponential law with one or two (if disc has
a break) exponential scales. Thus, the model of the brightness profile depends
on eight parameters (since $i$ and $PA$ of the galaxy are fixed): central surface
brightness of a bulge $\mu_b^0$, its \ser~ index $n$, and effective radius 
$r_e$, central surface brightnesses of inner and outer discs $\mu_{d_1}^0$, 
$\mu_{d_2}^0$, their exponential scales $h_1$ and $h_2$ and the location of the break $r_b$:

\begin{eqnarray}
I(r) = I_b^0  {e}^{-\nu_n \left( \frac{r}{r_e} \right)^{\frac{1}{n}}}+
\left\{ \begin{array}{l} I_{d_1}^0  {e}^{-\frac{r}{h_1}}, \; r < r_{b} \\
 I_{d_2}^0  {e}^{-\frac{r}{h_2}}, \; r>r_{b}\end{array}\right.,
\label{eq5}
\end{eqnarray}

where $I(r)$, $I_b^0$, $I_{d_1}^0$ and $I_{d_2}^0 $ are surface brightness
of a galaxy in intensities, 
and central intensities of a bulge, inner and outer discs ($I=10^{0.4( {const}-\mu)}$);
 $\nu_n$ is a constant depends on $n$. Despite the fact that the formula (\ref{eq5})
is written in terms of intensities, the fitting process is much simpler in terms of
surface brightnesses $\mu(r)$, $\mu_b^0$, $\mu_{d_1}^0$ and $\mu_{d_2}^0$, where the
exponential profiles are just straight lines and the range of values is
relatively small.
If the disc does not have a break, then $\mu_{d_1}^0 = \mu_{d_2}^0$ and
$h_1 = h_2$.

The number of free parameters can be reduced to seven: central surface brightness
of the outer disc $\mu_{d_2}$ can be expressed by the rest three parameters of the
discs in assumption that the brightness does not have discontinuities at the break point:
$\mu_{d_2}^0 = \mu_{d_1}^0+1.0857 r_b\left( \frac{1}{h_1} - \frac{1}{h_2}\right) $.

The fitting of the brightness profile $\mu(r)$ by the function (\ref{eq5}) was
made via {\small PYTHON}'s wrapper of {\small ODRPACK} (orthogonal distance regression)
{\small FORTRAN} library\footnote{http://www.scipy.org/doc/api\_docs/SciPy.odr.odrpack.html}.

\begin{figure}
\centering
\includegraphics[width=7.5cm, angle=0, clip=]{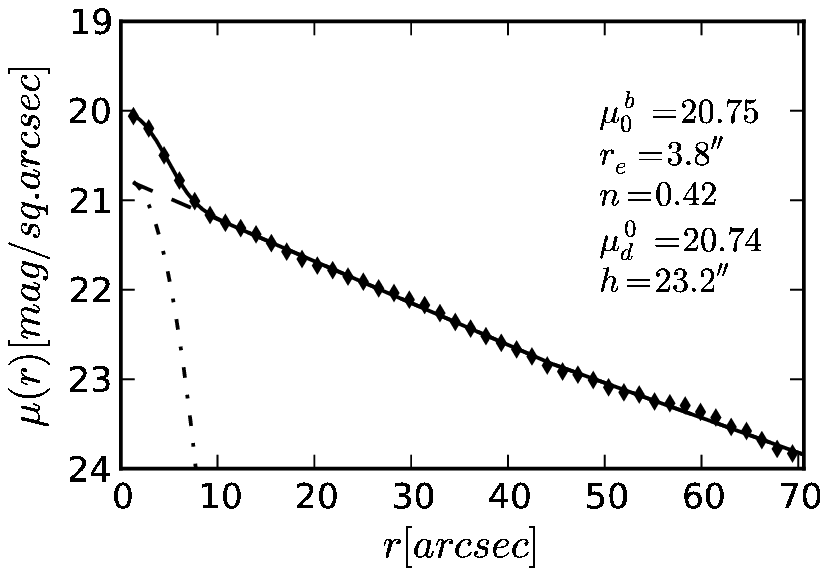}
\includegraphics[width=7.5cm, angle=0, clip=]{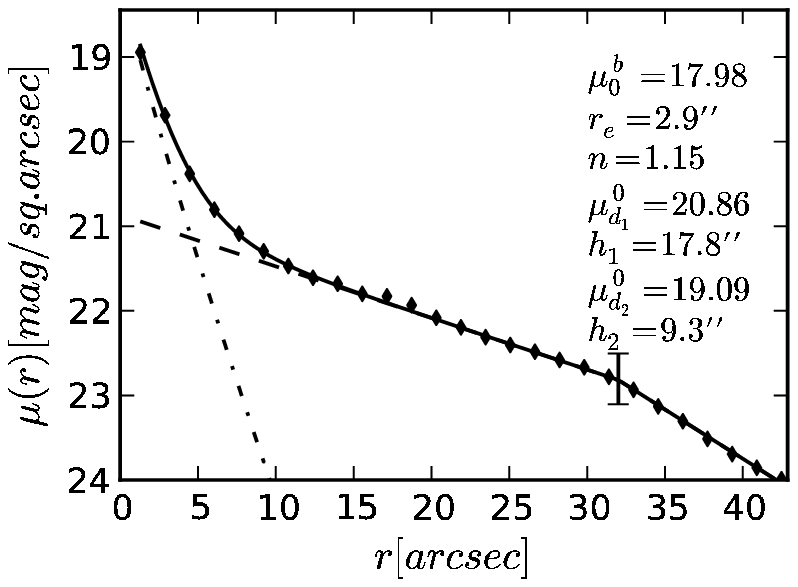}
\caption{The decomposition of two galaxies, PGC~38916 (top) and PGC~2182
(bottom) in g band. Diamonds -- observable distribution, dashed line -- disc model, 
dot--dashed line -- bulge, solid line -- the sum of the bulge and the disc.
Vertical dash on bottom picture at $r\approx 32''$ shows the break location.
Surface brightnesses are in magnitudes per square arc second, effective radius
and exponential scales -- in arc seconds.}
\label{fig2}
\end{figure}

The results of the decomposition of two galaxies without and with a break are
shown in Fig.~\ref{fig2}.

This decomposition allows us to obtain all eight parameters, described above, and
parameters that depend on them -- total bulge and disc luminosities, 
bulge-to-disc ratio etc.
At this stage we also measured the $g-r$ colour of disc.

One of the main benefits from a galaxy image decomposition is in
obtaining a high contrast image of spiral structure. The model (\ref{eq5}) 
does not account for the spiral
pattern, but only for large-scale components, so, after subtraction of this model
from the image of the galaxy, the bulge and the disc will fade away and spiral arms
will be visible on almost clear background.

\subsection{Pitch angles}

\begin{figure*}
\centering
\includegraphics[width=14cm, angle=0, clip, trim = 0mm 10mm 0mm 0mm]{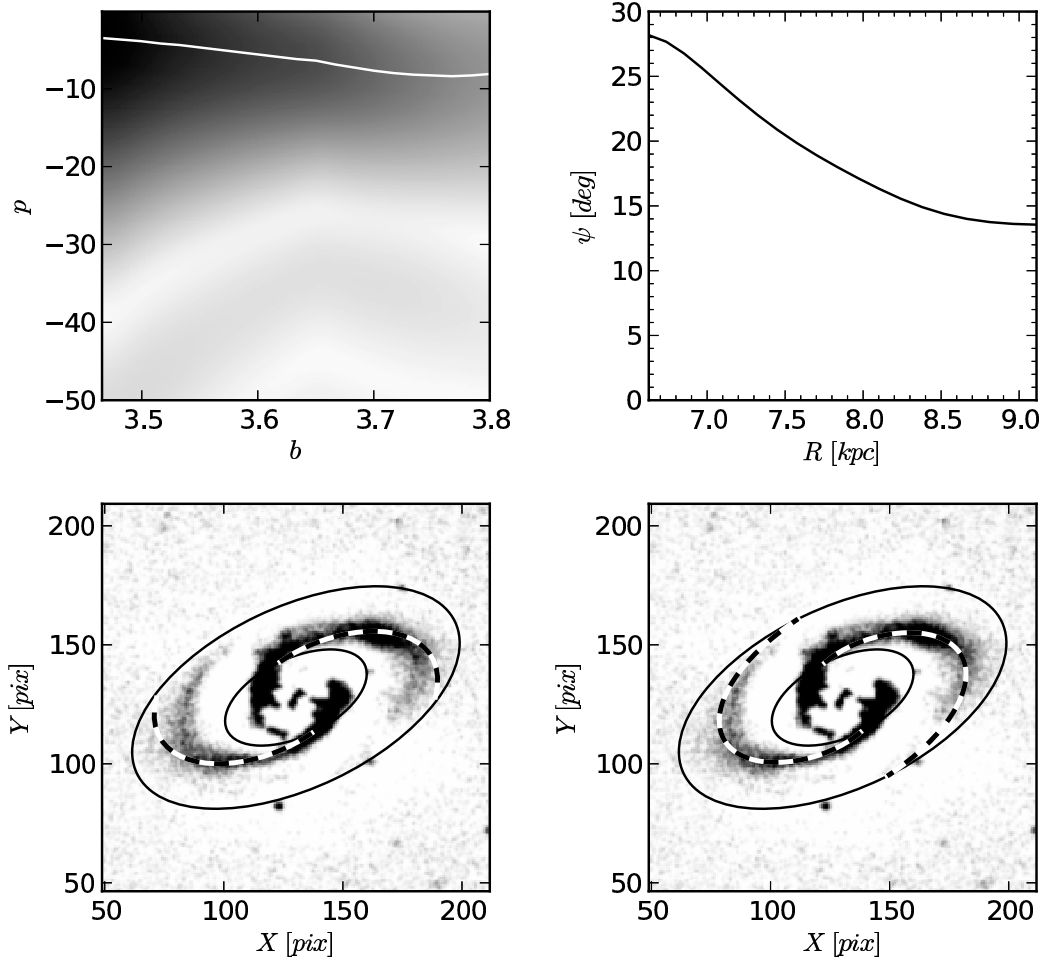}
\caption{The measurements of the pitch angle variations for PGC~22596.
Top left image shows two-dimensional Fourier transform. Darker colour shows
higher values of $|G(p,m,b)|$. Top right figure is the pitch angle value
as a function of a distance from the galactic centre. Bottom figures are
images of the galaxy (bulge and disc was subtracted) with overlapped model spiral
arms with the constant (left) and variable (right) pitch angle. Ellipses show
inner and outer radii of the spiral structure.}
\label{fig3}
\end{figure*}

The Fourier transform of a galaxy image is a widely used technique to obtain pitch angles 
of spiral arms (for example, \citealt{considere1982}, \citealt{davis2012}). 
It allows us to make decompositions
of a spiral pattern on to set of logarithmic spirals with different pitch
angles (just like the regular Fourier-decomposition of a time series on to sum of different
sinusoidal harmonics).

Common representation of this technique is as follows. Let $I(u, \Theta)$ be
the value of intensity in the point with log-polar coordinates $u\equiv\ln(r)$
and $\Theta$ (the origin of the coordinate system is in the centre of a galaxy), then

\begin{equation}
A(m, p) = \frac{1}{D} \int_{-\infty}^{\infty} \int_{-\pi}^{\pi} I(u, \Theta)  {e}^{- {i}(m\Theta+pu)} \,  {d}\Theta  {d}u,
\label{eq1}
\end{equation}

where $A(m, p)$ is a value of the contribution of the $m$-armed logarithmic spiral with
the pitch angle $\psi=\arctan\left(-\frac{m}{p}\right)$ in the whole spiral pattern. $D$ -- is a
normalization factor:

\[
D = \int_{-\infty}^{\infty} \int_{-\pi}^{\pi} I(u, \Theta) \, d\Theta du.
\]

The resulting value of the pitch angle is the pitch angle of the spiral
with larger contribution in Fourier-decomposition:

\begin{equation}
\psi = \arctan\left(-\frac{m}{p_{max}}\right).
\label{eq2}
\end{equation}

This method allows us to compute only the average value of the pitch angle over the whole
spiral structure even if different regions of the spiral pattern have different
pitch angle values.

To recover the information about spatial variations of the pitch angle value,
one can apply this method to different regions of the spiral structure. 
This can be done by introducing a spatial filter which can ``cut'' a
limited ring-like area of galaxy, thus one can compute the Fourier 
transform~(\ref{eq1}) only for this limited area. Changing the position
of spatial filter gives the pitch angle value for different areas.

It is convenient to use simple rectangular window as such spatial filter:

\begin{equation}
W(a, b, u) = \left\{ \begin{array}{l} 1, u \in [b-\frac{a}{2}, b+\frac{a}{2}] \\ 0, u \notin [b-\frac{a}{2}, b+\frac{a}{2}]  \end{array}\right.,
\label{eq3}
\end{equation}
where $a$ is width of the window and $b$ its position.

The regions for the pitch angle measurements were specified for each galaxy
by hand on the basis of its spiral structure visibility: we have manually selected
two radii $u_{0}$ and $u_{1}$ between which the spiral structure is clearly
visible.

Finally we have

\begin{equation}
 G(p, m, b) = \frac{1}{D} \int_{u_{0}}^{u_{1}} \int_{-\pi}^\pi I(u, \Theta)  {e}^{- {i}(m\Theta + pu)} W(a,b,u) \,  {d}\Theta  {d}u.
 \label{eq4}
\end{equation}

$G(p, m , b)$ gives a contribution of $m$-armed logarithmic
spiral with the pitch angle $\psi=\arctan\left(-\frac{m}{p}\right)$
at the distance $b$ from the centre of a galaxy. Similar to
one-dimensional case, the pitch angle value at any particular
distance can be found as a value of the pitch angle of spiral arm, which
gives largest contribution in the decomposition at this distance
(equation \ref{eq2}).

Fig.~\ref{fig3} presents an example of this method application 
for galaxy PGC~22596. The figure consists of four subimages.
First subimage shows two-dimensional map of the windowed Fourier analysis 
(equation \ref{eq4}),
where $b$ is a logarithm of distance from the centre of the galaxy and 
$p=-\frac{m}{\tan{\psi}}$.
Darker colour indicates higher values of $G(p, m, b)$ and white line indicates the 
maximum position as a function of the distance in the figure. The second subimage shows
the pitch angle value as a function of distance from the centre of the galaxy. 
On the third subimage two logarithmic spiral arms with average pitch angle 
are drawn above the residual (after bulge and disc models subtraction) image of the 
galaxy. For the comparison, on the fourth image two spirals with variable 
pitch angle are shown. These spiral arms with variable pitch angle 
can be computed by this recurrent formula:
\[
\left\{ \begin{array}{l} u_{n+1} = u_{n} + \tan(\psi(u_n)) \Delta\phi \\ \phi_{n+1} = \phi_n + \Delta\phi  \end{array}\right. .
\]
It is easy to see that spiral arms with variable pitch angle fit spiral
pattern of the galaxy better than ordinary logarithmic spirals with constant pitch angle.

\section[]{Results}

\subsection{Decomposition results}

To check the reliability of our model to describe the global photometric 
structure of the studied galaxies (Sect. 3.2), we have calculated their total luminosities 
using the decomposition parameters from Table 1. 
The luminosities of bulges and discs were computed from their decomposition
parameters:
\[
\left. \begin{array}{l} 
L_b = \frac{2\pi n}{\nu_n^{2n}}\Gamma(2n)I_0 r_e^2  \\ 
L_d = \left\{ \begin{array}{l} 2\pi I_0 h^2 \, \mbox{(disc without a break)}\\
      2\pi[ I_{0,1} h_1^2(1-(1+\frac{r_b}{h_1}) {e}^{-\frac{r_b}{h_1}})+ \\
       \quad+I_{0,2}h_2^2(1+\frac{r_b}{h_2}) {e}^{-\frac{r_b}{h_2}}] \, \mbox{(disc with a break),}  \end{array}\right.
\end{array}\right.
\]

where $\Gamma(2n)$ is the gamma function and $I_0$ is the central
intensity $I_0 = 10^{0.4( {const}-\mu_0)}$.

We have found that the mean
difference of the SDSS \textit{deVMag} 
magnitudes and our model magnitudes in the $g$ filter is $+0.^m08 \pm 0.^m09$. 
In the rest of our discussion we will use our model magnitudes.

Only two galaxies from our sample (PGC~38916 and PGC~39038) have single-disc 
surface brightness profiles. 15 galaxies (30\%) have upbending profiles and 33 galaxies (64\%)
demonstrate downbending profiles. This statistics is in agreement with 
\cite{pohlen2006} results.

\begin{figure}
\centering
\includegraphics[width=7.0cm, angle=0, clip=]{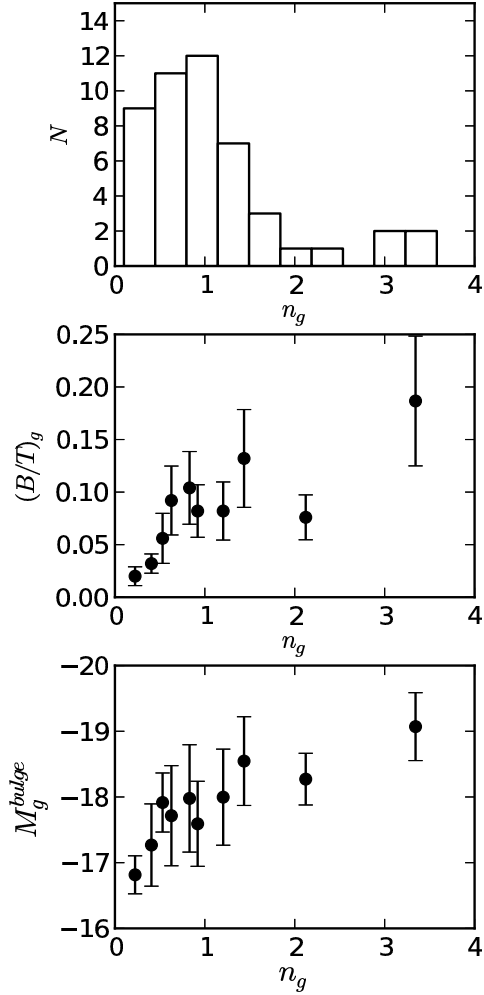}
\caption{Decomposition parameters of bulge: a distribution \ser~ indexes (top),
\ser~ indices versus bulge-to-total ratio (centre) and \ser~ index versus bulge
luminosity (bottom).}
\label{fig4}
\end{figure}

Fig. \ref{fig4} summarizes several known correlations concerning 
decomposition parameters. The top panel shows the distribution of the galaxies
by the \ser~ index $n$ of their bulges. Most of our galaxies are of 
late-types (Fig. 1), so values of their \ser~ index concentrate around 1 
(pseudo-bulges), although several classic bulges (small peak at $n\sim 3-4$) present too. 

The middle panel of Fig. \ref{fig4} shows correlation 
\ser~ index~--~bulge-to-total ratio. It is clear that galaxies with
larger bulge-to-total ratio have larger values of the \ser~ index. 
The bottom panel shows the \ser~ 
index~--~absolute magnitude of the bulge correlation. 
One can see that bulges with larger \ser~ index are brighter.

The correlations shown in Fig. \ref{fig4} have a significant scatter, 
so the points on the panels show average of five galaxies each. Error bars illustrate 
the scatter (rms) of real values in the corresponding bin. In the rest of the paper
we will use the same averaging in the figures.

The correlations on Fig. \ref{fig4} are typical for
spiral galaxies (e.g. \citealt{mosenkov2010} and references therein), thus one can conclude
that we have the sample of ordinary spiral galaxies.

\subsection{Pitch angles and their variations}

According to our measurements, the region of the spiral structure visibility (that is, where
the spiral arms can be clearly detected) spans, on average, 
from $0.74 \pm 0.09$ to $2.40 \pm 0.25$ of inner discs exponential scales ($h_1$).

For 4 of 5 galaxies with published estimates of their pitch angles we have
obtained very similar values: the mean difference between our estimates and the data
of other authors  is $-0.\degr2 \pm 0.\degr7$ (see Table 1).
The only galaxy with significant difference between our measurements and 
other estimates is PGC~23028. Kennicutt (1981) and Ma (2001) give for this galaxy 
the angles of 
13$\degr$ and 12.$\degr$2 respectively, while our value is 16.$\degr$7. We argue that 
this difference may be due to radial variations of the pitch angle: in the inner regions 
of this galaxy the value of the pitch angle is about 21$\degr$--22$\degr$ while in outer 
is only 13$\degr$--14$\degr$, which is close to values obtained by Ma and Kennicutt.

\begin{figure}
\centering
\includegraphics[width=6.0cm, height=5.8cm, angle=0, clip=]{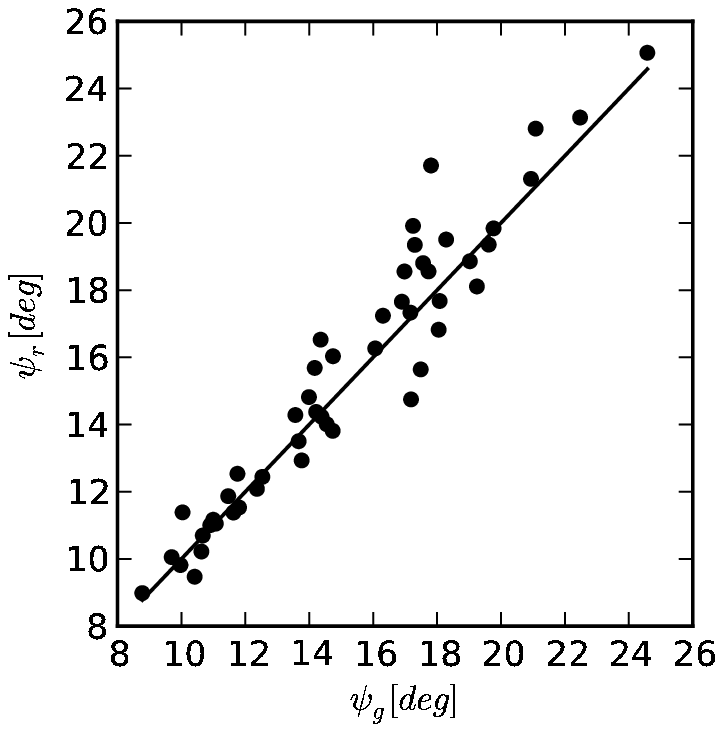}
\includegraphics[width=6.2cm, height=5.8cm, angle=0, clip=]{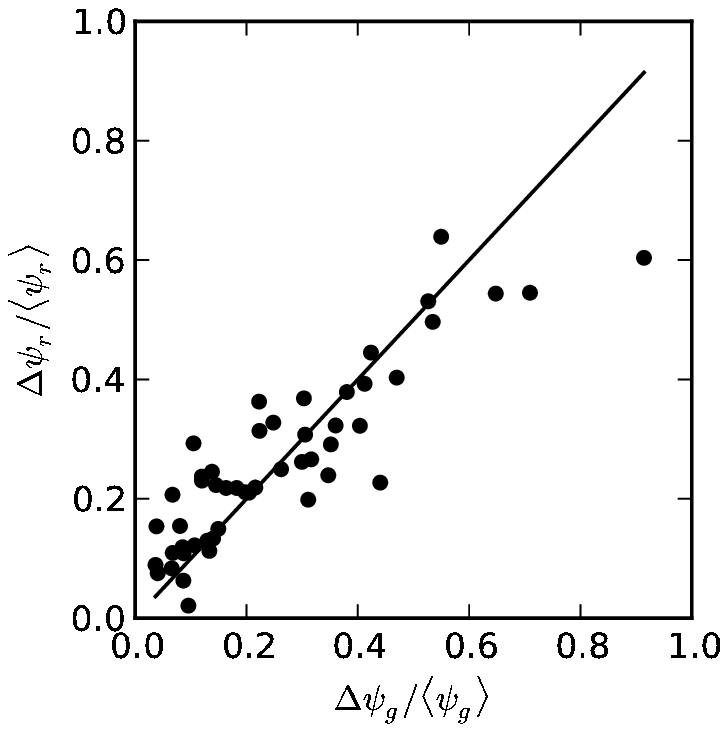}
\caption{Comparison of the average pitch angles (top panel) and its variations (bottom)
in $g$ and $r$ passbands. Solid lines show bisector $y=x$.}
\label{fig5}
\end{figure}

Fig. \ref{fig5} shows relations between pitch angles (top
panel) and amplitudes of their variations (bottom panel) measured
in different passbands. The amplitude of the pitch angle variation
is defined as maximum deviation from the average value divided by the average value.
The mean difference between pitch angles measured in different passbands
is $\overline{<\psi_g>-<\psi_r>}=-0.33\degr \pm 0.16\degr$ and the mean difference between
relative pitch angle variations in different passbands is 
$\overline{\frac{\Delta \psi_g}{<\psi_g>} - \frac{\Delta \psi_r}{<\psi_r>}} = -0.004 \pm 0.013$.

In most galaxies, the pitch angle decreases with increasing distance from the centre
(32 of 50 galaxies or 64\% $\pm$ 11\%), in the rest of galaxies the angle increases
up to maximum distance to which spiral arms are still visible.

\begin{figure}
\centering
\includegraphics[width=7cm, angle=0, clip=]{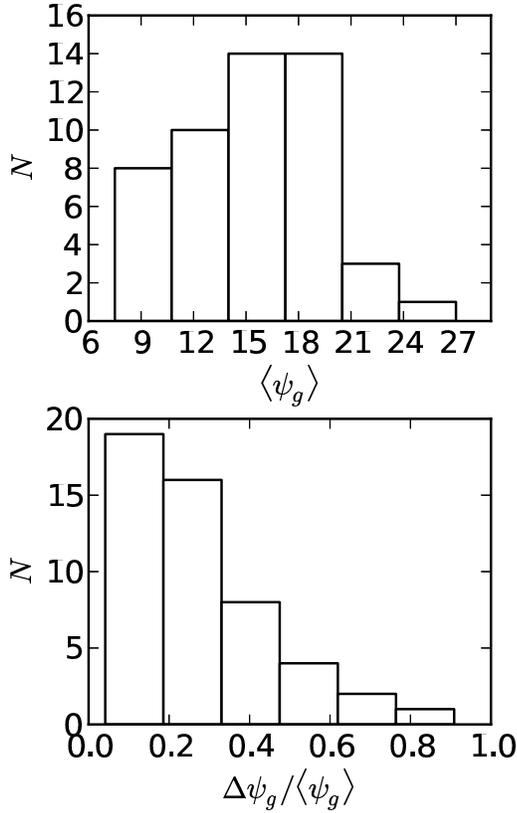}
\caption{Distributions of the sample galaxies over (top) the average values of 
pitch angle in degrees, (bottom) pitch angle variations in the $g$ filter.}
\label{fig6}
\end{figure}

Fig. \ref{fig6} presents the histogram illustrating the
distribution of the average values of pitch angle (top) and  the distribution of 
the sample galaxies by
the value of their pitch angle variation (bottom). 
As one can see, the maximum is around zero
and monotone decrease towards higher values of $\Delta\psi_g/\langle\psi_g\rangle$.
About 2/3 of galaxies have the variations of the pitch angle greater
than $0.2$. Only $\sim$1/10 of the sample galaxies shows almost constant
pitch angles with variations $< 10\%$.

\subsection{Pitch angles and general parameters of galaxies}

\begin{figure}
\centering
\includegraphics[width=7cm, angle=0, clip=]{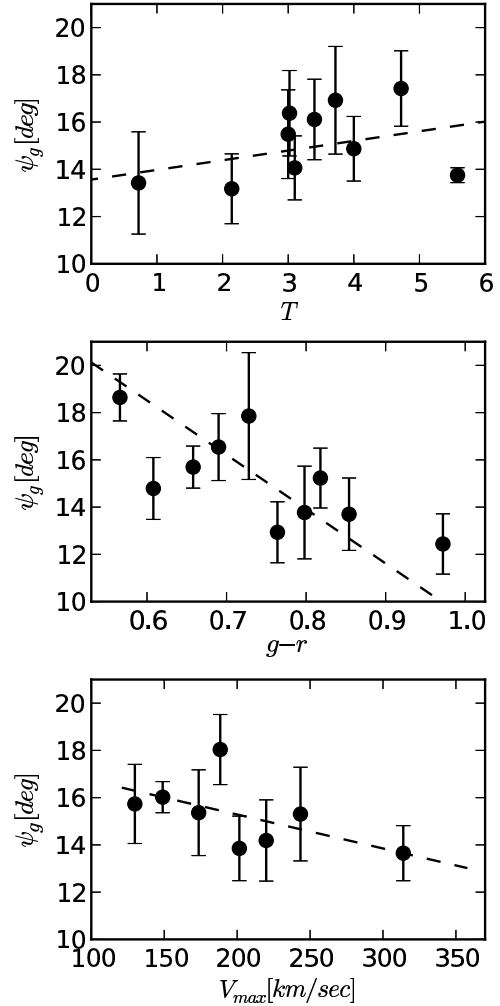}
\caption{From top to bottom: the average pitch angles in the $g$ passband 
as a function of the morphological type, of the $g-r$ galaxy colour, of the maximum 
rotation velocity. The dashed lines correspond to the regression lines.}
\label{fig7}
\end{figure}

Fig. \ref{fig7} shows several known observational trends of
the average pitch angles with general parameters of galaxies.
As one can see in this figure, early-type, red and massive (with
large values of $V_{max}$) galaxies, on average, demonstrate tighter
spiral arms in comparison with late-type, blue and less massive 
(see also \citealt{kennicutt1981}, \citealt{ma2002}, \citealt{sr2011}).

We found no statistically significant correlation between the pitch angles 
and the total luminosity of galaxies. We also tested the possible dependence 
of the pitch angle values on the ratio of the dynamical mass to the stellar mass
within four scalelengths $h_1$ (e.g. \citealt{zasov2002}, \citealt{kregel2005}).
This correlation is insignificant also, although the size of our sample is not
enough large for definitive judgement.

\subsection{Pitch angles and galactic bulges and discs}

Here we summarize some relations between the average pitch angles and 
parameters of galactic bulges and discs. There are
seven parameters in the photometric model and some of them show a notable
correlation with the pitch angles.

\begin{figure}
\centering
\includegraphics[width=8.5cm, angle=0, clip=]{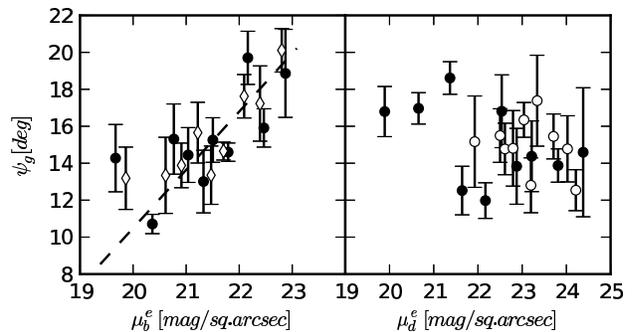}
\caption{Pitch angle versus effective brightness of bulge (left) and disc (right).
Open rhombs in the left-hand figure show bulges with $n \leq 2$.
On the right-hand side open circles represent values for inner discs, black
ones are for outer discs.}
\label{fig8}
\end{figure}

Fig. \ref{fig8} shows the correlation of the pitch angle
with effective surface brightnesses both of the bulge and the disc.
Although during the decomposition process we have measured the central
surface brightnesses, for this figure we have recomputed
them to effective surface brightnesses because for bulges with $n>1$ the 
extrapolation of the model light curve to the centre of the galaxy gives 
too high values of the central surface brightness, whereas effective surface brightness
takes more moderate values. The effective surface brightness for given
central surface brightness and the \ser~ index $n$ can be found as
$\mu^e = \mu^0 + 2.5\nu_n / \ln 10$.

Fig. \ref{fig8} demonstrates that there is a statistically
significant correlation between the effective brightness of the bulge and
the pitch angle of spiral arms: galaxies with brighter bulge have
smaller pitch angles. The correlation looks somewhat better for pseudo-bulges
with $n \le 2$ (open rhombs in the figure).
However there is no statistical significance
for $\psi - \mu^e$ relation for discs (right-hand panel of Fig. \ref{fig8}).

\begin{figure}
\centering
\includegraphics[width=9cm, angle=0, clip=]{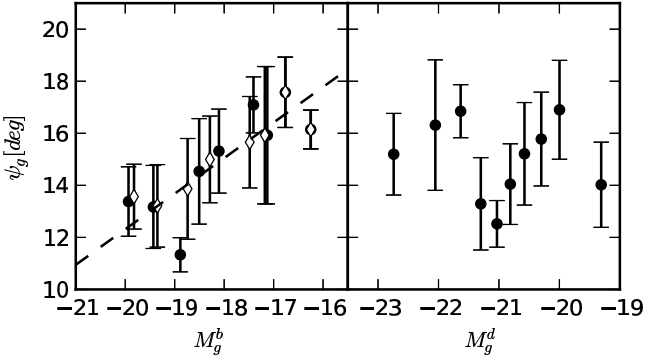}
\caption{Pitch angle versus absolute magnitude in the $g$ band of 
bulge (left) and disc (right). Open rhombs in the left-hand figure show bulges 
with $n \leq 2$.}
\label{fig9}
\end{figure}

The pitch angle--absolute magnitude relation (Fig. \ref{fig9})
shows the same behaviour. There is a significant correlation 
between the pitch angle and the absolute magnitude of bulge, 
whereas for galactic discs the correlation is absent. (Conversion from apparent magnitude 
to absolute one was made using luminosity distance and absorption in the Galaxy according 
to the NED database.) On the other hand, the correlation between average $\psi$ value
and effective radius of bulge is weak or absent.

\begin{figure}
\centering
\includegraphics[width=8.5cm, angle=0, clip=]{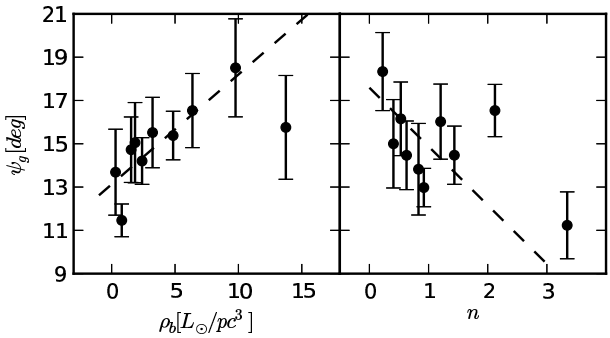}
\caption{Pitch angle versus the mean luminosity density of bulge (left) and \ser~
index (right) in the $g$ passband.}
\label{fig10}
\end{figure}

On the left-hand side of Fig. \ref{fig10} a correlation of the pitch 
angle and the luminosity density of the bulge is shown.
The luminosity density of the bulge is defined as luminosity of the
bulge (in solar units) per cubic parsec: 
$\rho_b = \pi L_b(\le r_e) / r_e^3 = 0.5 \pi L_b/r_e^3$,
where $L_b(\le r_e)$ is luminosity of the bulge inside its effective radius and
it is equal to a half of its total luminosity $L_b$.
Right-hand side of Fig. \ref{fig10} presents $\phi_g$ --
\ser~ index of the bulge relation. As one can see, there is clear
observational trend of the pitch angle with the bulge characteristics.

\subsection{Pitch angle variations and disc parameters}

\begin{figure}
\centering
\includegraphics[width=9cm, angle=0, clip=]{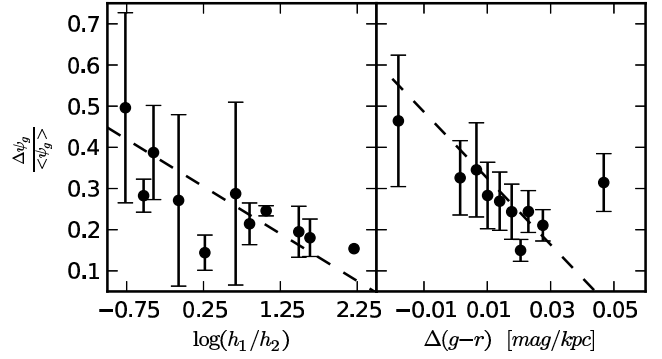}
\caption{Pitch angle variations as a function of disc parameters: variation
versus value of the break in a disc (left-hand panel) and variation versus $g-r$ colour
gradient.}
\label{fig11}
\end{figure}

Fig. \ref{fig11} displays dependence of pitch angle variation
on parameters of disc. The left-hand panel of Fig. \ref{fig11} 
shows how the relative variation of the pitch angle depends on
the value of the break in galaxy disc. Values on the
$x$-axis is a decimal logarithm of ratio of the inner exponential scale to
the outer one, so, if $\log(h_1/h_2)<0$, galaxy
has upbending brightness profile, $\log(h_1/h_2)>0$ means downbending or
truncated profile.

The right-hand panel of Fig. \ref{fig11} shows the pitch angle
variation versus $g-r$ colour gradient. We obtain the colour gradient
as the difference of disc colour at the beginning and at the end of the spiral
structure (which was found during the decomposition stage) divided
by the distance between them.

Both panels demonstrate notable observational trends. The first one is that  
stronger pitch angle variations associated with envelope-type surface brightness 
distributions ($h_1 < h_2$), while galaxies with almost constant pitch angles
are more frequent among truncated discs ($h_1 > h_2$). Also, galaxies with
positive colour gradient (red in the centre, blue in the periphery) show
relatively small variations, while flat or negative gradients are associated
with large variations of pitch angles.

\subsection{Pitch angle variations and the galaxy environment}

Although we have selected non-interacting galaxies only (by setting
the \textit{multiplicity} parameter in the EFIGI to be equal zero), it does not mean
that galaxies of our sample does not have any satellites at all.

Nearby companions can distort or even induce (e.g. \citealt{dobbs2010}) spiral
structure. To check if variations of the pitch angle of our sample galaxies 
can be related with their environment, we have compared the mean number of
satellites for 10 galaxies with minimal pitch angle variations
and for 10 more with maximal variations.

The searching for satellites was made by examining the area around
the galaxy inside some fixed radius and in depth by some fixed magnitude.
We have made the search with two different sets of parameters:
the first is $R_{ {search}}=10 R_{ {petro}}$, $\Delta m \leq 2^m$ and the
second one $R_{ {search}} = 20 R_{ {petro}}$, $\Delta m \leq 3^m$, where $R_{ {petro}}$
is the Petrosian radius of the galaxy according to SDSS, and $\Delta m$ is
the difference between apparent magnitudes of the galaxy and its companion. 
In the second
case many background galaxies fall in the searching area, but the
true satellites were separated from them by the redshift value.

The results of the companions searching are shown in the Table 2.

\begin{table}
\centering
\begin{tabular}{cccc}
  \hline
  Subsample & $\left< \frac{\Delta \psi}{<\psi>}  \right>$ & $<n_1>$ &   $<n_2>$   \\
     (1)    &              (2)                             &   (3)   &   (4)      \\
  \hline
  $\min\frac{\Delta \psi}{<\psi>} $ & $0.09 \pm 0.02$ & $0.5\pm0.9$ & $1.9\pm2.1$ \\
  $\max\frac{\Delta \psi}{<\psi>} $ & $0.59 \pm 0.14$ & $0.3\pm0.5$ & $1.7\pm1.8$ \\
 \hline
\end{tabular}
 \caption{The pitch angles variations and the environment of the galaxies.}
 \label{tabSatellites}
\end{table}

The columns of this table are: (1) a subsample (10 galaxies with
smallest or 10 galaxies with largest pitch angle variations); (2) the
average pitch angle variation for this subsample; (3) the mean number of
satellites for the first set of searching criteria ($10 R_{ {petro}} $ and $2^m$) and
(4) the mean number of satellites for the second set of searching
criteria ($20 R_{ {petro}}$ and $3^m$).

It is clear, that the pitch angle variation does not depend on
the number of satellites at least for our sample of non-interacting
galaxies. We argue that it means that the pitch angle variations
of these galaxies are generated by their intrinsic properties but
not by tidal perturbations with nearby galaxies.

\section[]{Conclusions}

We have presented a detailed photometric study and pitch
angle measurements for a sample of 50 non-barred or weakly barred
Sa--Sc galaxies with two large-scale spiral arms. 
The main results of this work are as follows.

\begin{enumerate}

\item We have developed and described a new method for measuring the pitch
angle ($\psi$) of the spiral arms of galaxies based on the window Fourier analysis. 
This method allows not only to infer the average pitch angle, but also to obtain 
its value as a function of galactocentric radius. 

\item Application of this technique for a sample 
of 50 galaxies showed that, in general, spiral arms of most 
galaxies in the sample cannot be described by a single value of the pitch angle.
About 2/3 of galaxies demonstrate pitch angle variations exceeding 20\%.

\item We have found that pitch angle variations does not depend on the presence 
of close companions -- galaxies with large and small pitch angle variations are 
in approximately the same local spatial environment. From the other side, 
variations of $\psi$ correlate with the properties of galaxies themselves --
with the shape of the surface brightness distribution, and with the sign of
stellar disc colour gradient (Fig. 11).

\item Average pitch angle demonstrates known dependencies on general characteristics
of galaxies -- early-type, red and massive galaxies tend to have 
tighter spiral arms in comparison with late-type, blue and less massive (Fig. 7).

\item Average pitch angle shows clear observational trend with general
properties of galactic bulges -- faint bulges tend to have 
opened spiral arms, while bright (with high surface brightness) and luminous
(with large total luminosity and large \ser~ indices) bulges
demonstrate tight spiral arms (Figs. 8--10). On the other hand, correlation 
of $\psi$ with disc parameters (Figs. 8 and 9) is much weaker. Therefore,
classic correlation between the pitch angle of a galaxy and its morphological
type can be explained by changes in the properties of bulges along the Hubble
sequence.

\end{enumerate}

Our main conclusion -- dependence of the spiral structure on the properties
of bulges (or central mass concentrations) -- is in according with expectations 
of two presently most pursued models for the formation of spiral structures
(density waves and manifold) (see discussion in \citealt{berr2013}).
The mass of the bulge of the galaxy correlates with the mass of the central black
hole (e.g., \citealt{mag1998}), and, therefore, there is a relationship 
between the pitch angle and the mass of the black hole (\citealt{seigar2008},
\citealt{berr2013}). As we know, neither theory gives clear predictions 
on the radial pitch angle variations. Therefore, any model that intends 
to explain the formation and evolution of the spiral pattern in disc galaxies
has to reproduce our empirical findings as well.

\section*{Acknowledgements}

We acknowledges partial financial support from the RFBR grant 11-02-00471a.

Funding for SDSS-III has been provided by the Alfred P. Sloan Foundation, the Participating Institutions, the National Science Foundation, and the U.S. Department of Energy Office of Science. The SDSS-III web site is http://www.sdss3.org/.

SDSS-III is managed by the Astrophysical Research Consortium for the Participating Institutions of the SDSS-III Collaboration including the University of Arizona, the Brazilian Participation Group, Brookhaven National Laboratory, Carnegie Mellon University, University of Florida, the French Participation Group, the German Participation Group, Harvard University, the Instituto de Astrofisica de Canarias, the Michigan State/Notre Dame/JINA Participation Group, Johns Hopkins University, Lawrence Berkeley National Laboratory, Max Planck Institute for Astrophysics, Max Planck Institute for Extraterrestrial Physics, New Mexico State University, New York University, Ohio State University, Pennsylvania State University, University of Portsmouth, Princeton University, the Spanish Participation Group, University of Tokyo, University of Utah, Vanderbilt University, University of Virginia, University of Washington, and Yale University. 

 This research has made use of the NASA/IPAC Extragalactic Database (NED)
 which is operated by the Jet Propulsion Laboratory, California Institute
 of Technology, under contract with the National Aeronautics and Space
 Administration. We acknowledge the usage of the HyperLeda data base (http://leda.univ-lyon1.fr)

The authors thank an anonymous referee for constructive comments and suggestions.
\bibliographystyle{mn2e}
\bibliography{art}

\label{lastpage}

\end{document}